\documentclass[aps,prl,twocolumn,groupadress]{revtex4}
\usepackage{graphicx}
\usepackage{amsmath}
\usepackage{amssymb}
\usepackage{epsfig}

\begin{document}

\title{Temperature dependence and resonance effects in Raman scattering
    of phonons in NdFeAsO$_{x}$F$_{1-x}$ single crystals}

\author{Y. Gallais$^{1}$}
\email{yann.gallais@univ-paris-diderot.fr}
\author{A. Sacuto$^{1}$}
\author{M. Cazayous$^{1}$}
\author{P. Cheng$^{2}$}
\author{L. Fang$^{2}$}
\author{H. H. Wen$^{2}$}

\address{$^1$ MPQ CNRS UMR 7162 Universit\'{e} Paris 7-Paris Diderot, B\^{a}timent Condorcet, 75205 Paris Cedex 13, France \\
    $^2$Institute of Physics, Chinese Academy of Sciences, Beijing 100190, China}
    
\begin{abstract}
We report plane-polarized Raman scattering spectra of iron oxypnictide superconductor NdFeAsO$_{1-x}$F$_x$ single crystals with varying fluorine $x$ content. The spectra exhibit sharp and symmetrical phonon lines with a weak dependence on fluorine doping $x$. The temperature dependence does not show any phonon anomaly at the superconducting transition. The Fe related phonon intensity shows a strong resonant enhancement below 2~eV. We associate the resonant enhancement to the presence of an interband transition around 2~eV observed in optical conductivity. Our results point to a rather weak coupling between Raman-active phonons and electronic excitations in iron oxypnictides superconductors.
\end{abstract}

\maketitle

\par
The discovery of a new class of high temperature superconductors belonging to the family of iron oxypnictides raises the possibilty of a new route to high T$_c$ superconductivity besides the one of the cuprates \cite{Kamihara08}. Most of the compounds discovered up to now have the formula ReFeAsO$_{1-x}$F$_x$ where Re is rare-earth atom (La, Nd, Sm). Superconductivity is believe to take place in the conducting FeAs layer where the Fe and As atoms are tetrahedrally coordinated and the Fe atoms form a two-dimensionnal square lattice. The chemical substitution of O with fluorine F allows electron doping of the FeAs planes and increases T$_c$ up to 55~K in SmFeAsO$_{0.9}$F$_{0.1}$ \cite{Ren}. 

\par
While there are similarities with the cuprates, two-dimensionality and the close proximity of high temperature superconductivity with magnetic order for example, it is becoming increasingly clear that these compounds also bear significant differences with them: the undoped compound is not a Mott insulator but a bad metal with antiferromagnetic spin density wave (SDW) order arising from the Fe moments \cite{Cruz,Chen-neutrons}. The SDW order is associated with a structural transition \cite{Cruz,Qiu,Chen-neutrons} and both the magnetic order and the structural phase transition are suppressed with fluorine doping, leading to the emergence of superconductivity \cite{Cruz,Qiu}. The role of doping in these compounds appears to be quite different than in the cuprates. In iron oxypnictides, doping may in fact be just one of the different ways of tuning and/or suppressing various competing itinerant magnetic orders \cite{Mazin-Johannes}. For example pressure was already shown to be an alternative tuning parameter \cite{Park,Alireza}.

\par
 Whether superconductivity in these compounds arises from strong electronic correlations as is believed in the cuprates, from a more conventional phononic mechanism or from a completely different route, remains an open question. A peculiarity of the oxypnictides is the interplay between strong repulsion due to the localized chararacter of the Fe-3d bands and the orbital degrees of freedom \cite{Haule}. This interplay may provide a completely new mechanism of strongly correlated high temperature superconductivity.
\par
 Determination of the order parameter symmetry is crucial to understand the nature of the superconducting mechanism but measurements of the gap anisotropy are still very preliminary. The multiband nature of these compounds which arises from the various Fe related bands crossing the Fermi level complicates further the determination of the order parameter symmetry \cite{Lebesgue,Liu}. On the other hand, determination of the phonon dynamics may shed light on the possible role of phononic degrees of freedom in the superconductivity of oxypnictides. Ab-initio calculations give an electron-phonon coupling constant seemingly too weak to explain high T$_c$ superconductivity in these compounds \cite{Boeri}. Nevertheless recent reports of a nodeless gap via ARPES \cite{Ding,Kondo} 
and the strong sensitvity of the calculated band structures near the Fermi level to the distorsion the FeAs tetrahedra indicate the possible role of vibrational degrees of freedom  \cite{Mazin-2, Eschrig,Boeri,Singh,Georges}.Therefore the role of phonons and their coupling to electronic degrees of freedom deserved to be scrutinized.
  
\par
In this communication we report plane-polarized Raman scattering measurements of zone-center phonons in NdFeAsO$_{1-x}$F$_x$ single crystals. The crystals have a superconducting T$_c$ of about 50~K for $x$=0.18 and 48~K for $x$=0.30. The undoped crystals $x$=0 are non-superconducting. The single crystals have been grown by a flux method at ambiant pressure and have a nominal composition of x=0, x=0.18 and x=0.30 respectively \cite{Jia}. The typical lateral size of the crystals studied here is about 20x20$\mu$m.

\par
The measurements were performed using a micro-Raman set-up in back scattering geometry. Several excitation lines of an Ar/Kr laser were used ranging from 1.9~eV to 2.7~eV. The scattered light was collected and analyzed by a triple grating spectrometer (JY T64000) and a back-illuminated nitrogen cooled CCD camera. The room temperature measurements were performed with a x50 objective. For the low temperature measurement the crystals were mounted on the cold finger of cryostat and a long working distance x50 objective was used. 
\par
Special care was taken in order to avoid overheating the crystals and power densities were kept below 10$^3$~W/cm$^2$. The local heating at the laser spot was estimated by comparing both the temperature and the power dependences of the Raman spectra. For most of the measurements reported here the heating was estimated to be 30~K except for the lowest temperature measurement where a lower power density was used yielding an estimated heating of 20~K. All the temperatures displayed have been corrected for the laser heating. Since the crystals have plate-like shapes, only the scattering configuration in which polarizations are in the $ab$ plane could be measured. All the spectra were corrected for the spectral response of the spectrometer and the CCD detector. They were not corrected for the optical properties of NdFeAsO$_{1-x}$F$_x$.

\begin{figure}
\centering \epsfig{figure=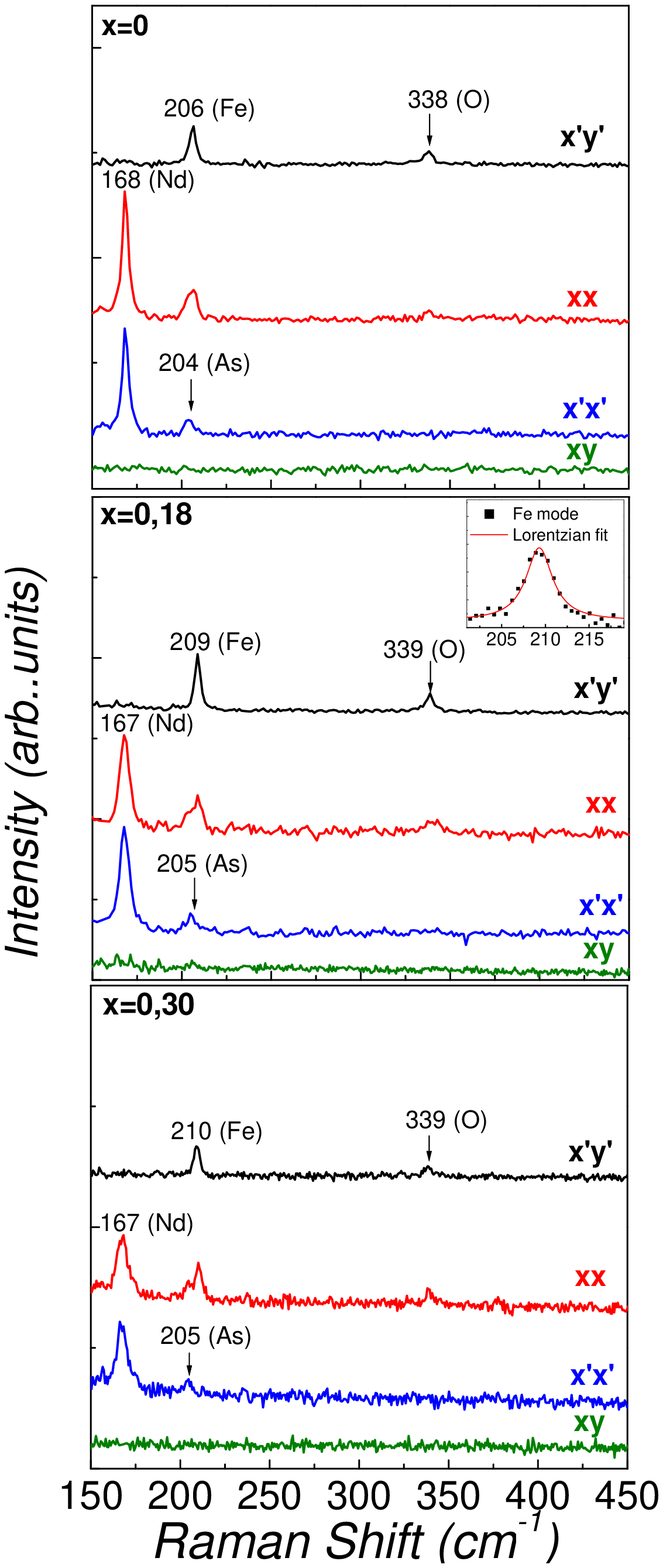, width=0.85\linewidth,clip=}
\caption{(Color online) Room temperature Raman spectra of NdFeAsO$_{1-x}$F$_x$ with incident and scattered light polarized in the $ab$ plane for three different fluorine concentrations (x=0, x=0.18 and x=0.30). The polarization configuration of each spectra is indicated in Porto notation. The frequency of each phonon and the atom involved are indicated. The inset shows a Lorentzian fit of the Fe mode for x=0.18 showing a half width at half maximum (HWHM) of about 2~cm${-1}$ } \label{fig1}
\end{figure}

\par
Figure \ref{fig1} shows the Raman spectra in different polarization configurations for x=0, x=0.18 and x
=0.30 using $\lambda_{exc}$=514,52~nm (2.4~eV). The Porto notation is used: the first letters refer to the direction of the incoming polarization with respect to crystallographic directions while the second one refers to the direction of the outgoing polarization. $x$ and $y$ refer to [100] and [010] directions respectively while $x'$ and $y'$ refer to [110] and [1-10] directions respectively. NdFeAsO$_{1-x}$F$_x$ has tetragonal symmetry at room temperature and fluorine doping is believe to supress the orthorombic distorsion that occurs at low temperature in the undoped compound  \cite{Cruz, Qiu}. The $xy$ configuration probes the B$_{2g}$ symmetry, the $x'y'$ B$_{1g}$ symmetry while the $xx$ and $x'x'$ probe the A$_{1g}$+B$_{1g}$ and A$_{1g}$ +B$_{2g}$ symmetries respectively.
\par
In agreement with the data and analysis of Hadjiev et al. for undoped SmFeAsO \cite{Hadjiev}, four zone centered Raman active phonons are found. Two have B$_{1g}$ symmetry and two have A$_{1g}$ symmetry. Their assignement was reported by Hadjiev et al. \cite {Hadjiev}:  the A$_{1g}$ modes at 167~cm$^{-1}$ and 205~cm$^{-1}$ arise from the out of plane motions of the Nd and As atoms respectively and the B$_{1g}$ modes at 209~cm$^{-1}$ (206~cm$^{-1}$ for x=0 and 210~cm$^{-1}$ for x=0.30) and 339~cm$^{-1}$ arise from the out of plane motions of the Fe and O atoms respectively. Except for the Fe mode the phonons frequencies show only weak changes with varying x content. The hardening of the Fe mode with increasing $x$ is consistent with the data of Le Tacon et al. which also show a hardening of this mode between x=0 and x=0.15 \cite{Tacon}.

\par
\begin{figure}
\centering \epsfig{figure=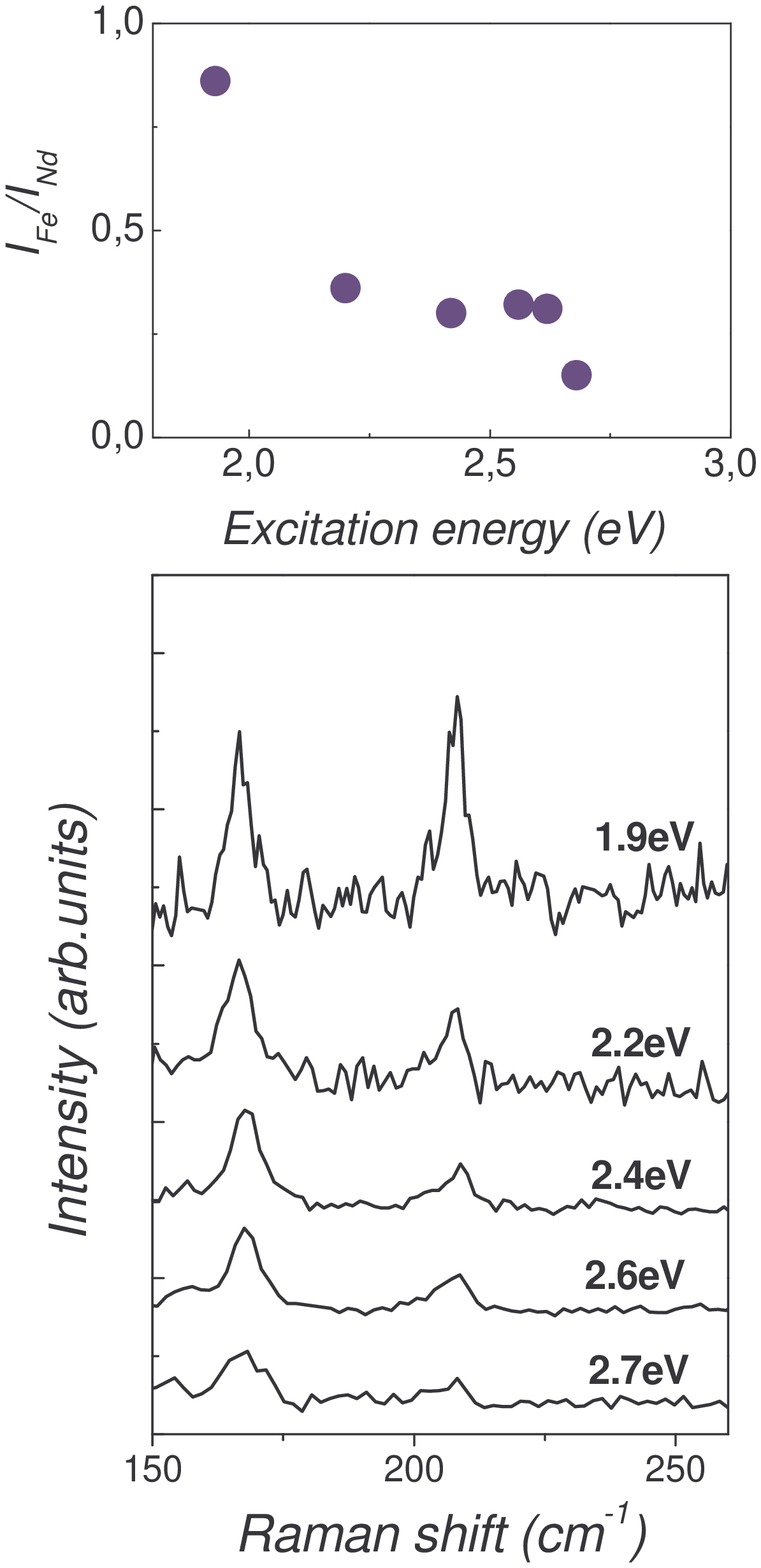, width=0.69\linewidth,clip=}
\caption{(Color online) Lower panel: room temperature Raman spectra as a function of laser excitation energy for the $xx$ polarization configuration and x=0.18. The upper panel shows the resonance profile of the Fe mode integrated intensity with respect to the Nd mode integrated intensity.} \label{fig2}
\end{figure}

An important aspect of the data reported here is that all the phonons are sharp and symmetrical. In particular the half width at half maximum (HWHM) of the Fe mode is only about 2 cm$^{-1}$ at room temperature for all doping x (see the inset of Fig. \ref{fig2}). This value is considerably lower than in the cuprates where the HWHM of most of the phonon lines at room temperature is at least 5~cm${-1}$. In addition, in the cuprates, several phonons lines exhibit strong coupling with the electronic continuum and show a distinctive asymmetrical Fano lineshape \cite{Cooper}. Correspondingly, except for a broadening of the Nd phonon line upon increasing x, fluorine doping has a very limited impact on the phonons lineshapes (linewidths and positions), again in contrast with cuprates like Y-123, or to a lesser extent Bi-2212, where phonon lines shows large renormalizations upon doping \cite{Altendorf,Martin,Opel,Hewitt}. Altogether, these observations suggest a rather weak coupling between in-plane polarized Raman-active phonons and electronic degrees of freedom in oxypnictides. We note that a strong coupling of the Fe in-plane breathing mode to the electronic continuum has been suggested \cite{Eschrig} but this mode is only accessible using polarization along the c-axis and is therefore not reported here.

\par

\par

In Fig. \ref{fig2} is displayed the Raman spectrum in $xx$ configuration for x=0.18 at room temperature as a function of the incident laser energy. Compared to the Nd mode, the Fe mode shows a significant increase in intensity towards when excited using near infra-red excitation energy (1.9~eV). When normalized to the integrated intensity of the Nd mode, its integrated intensity increases by almost a factor of 3 between 2,2~eV and 1.9~eV. Such a resonant profile of the Fe mode suggests the presence of an interband transition located around or below 2~eV involving a Fe-3d related band. Recent DMFT calculations indeed predicts an interband transition between As-4p and Fe-3d related bands around 2~eV \cite{Haule}. A recent optical study  of LaFeAsOF also found a weak interband transition in the same energy range \cite{Drechsler}. The reported resonance enhancement may provide a route to explore electronic Raman scattering from quasiparticle excitations arising from Fe-3d related Fermi surfaces.
\par

\begin{figure}
\centering \epsfig{figure=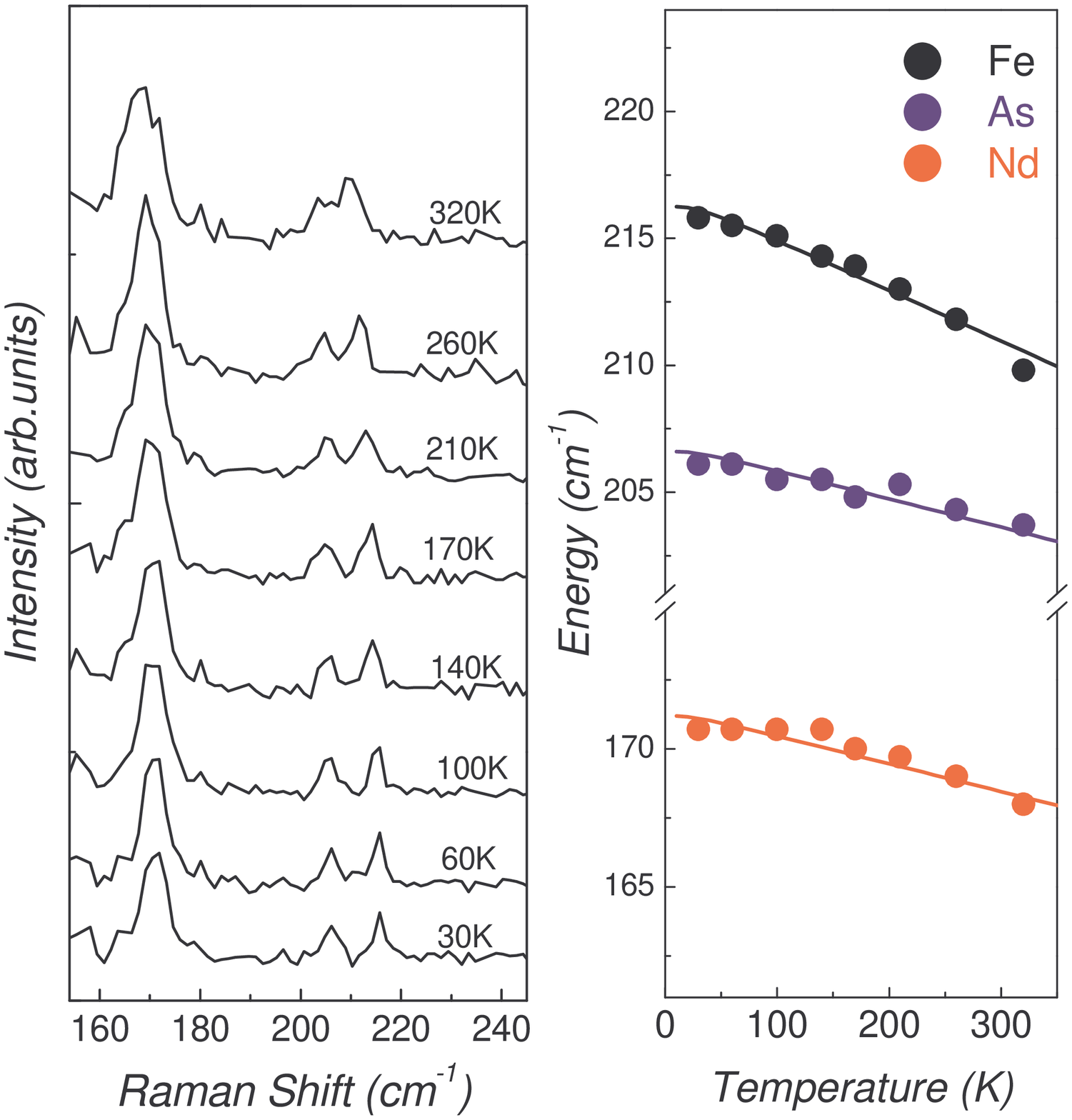, width=0.95\linewidth,clip=}
\caption{(Color online) Left panel: Temperature dependance of the Raman spectra in $xx$ polarization configuration for x=0.18. The evolution of the phonon frequencies as a function of temperature is reported in the right panel. The solid lines show a fit with a standard anharmonic behavior using the following expression for the phonon frequency: 
$\omega$(T)=$\omega_0$ 
+ C(1+$\frac{2}{\exp{\frac{\hbar\omega_0}{2k_BT}}})$ where $\omega_0$ is the bare phonon frequency and $C$ a temperature independent fitting parameter \cite{Klemens}.
}
 \label{fig3}
\end{figure}

The temperature dependence of the Raman spectrum between 320~K and 30~K for x=0.18 is shown in Fig. \ref{fig3}. The Nd, Fe and As modes frequencies show modest temperature dependences. The Fe mode shows a slightly more pronounced hardening upon cooling. This results in an enhanced splitting of the As and Fe modes which become clearly resolved upon lowering temperature. No anomaly in linewidths and positions is detected when crossing T$_c$ for all three modes. The same behavior was reported by Litvinchuk et al. in related, oxygen-free K$_{x}$Sr$_{1-x}$Fe$_2$As$_2$ single crystals \cite{Litvinchuk}. This is in sharp contrast with the cuprates where anomalies in phonon frequencies and linewidths are often observed upon crossing T$_c$ \cite{McFarlane,Zeyher,Cardona,Altendorf,Hadjiev-2,Zhou,Hewitt}. In the entire temperature range, the phonon frequencies can be reproduced by a standard anharmonic decay model \cite{Klemens} in which a zone-centered optical phonon decays into two lower-frequency acoustical phonons. The result of the fit is shown in fig. \ref{fig3} for the Nd, As and Fe modes. Again the absence of phonon anomalies at T$_c$ highlights the absence of a significant coupling between the Raman-active phonons and the electronic degrees of freedom.

\par   

In conclusion we have reported plane-polarized Raman spectra of NdFeAsO$_{1-x}$F$_x$. The doping dependence and the temperature dependence of the phonon modes suggest a weak coupling between Raman-active modes and electronic excitations. The situtation is in contrast with the cuprates where several phonons line exhibits strong lineshape renormalizations due to electron-phonon coupling. This weak electron-phonon coupling may prevent the use of phonons to gain insight into the electronic degrees of freedom as was done in the cuprates \cite{Cardona}. The direct observation of electronic Raman scattering will likely requires larger single crystals but the strong enhancement of the Fe phonon mode intensity below 2~eV reported here suggests the possibility of resonantly enhancing the electronic Raman scattering cross-section. This may shed light into the symmetry of the superconducting order parameter in iron oxypnictides.

\par


\end{document}